\documentclass[twocolumn,aps,prb,showpacs,amsmath,amssymb]{revtex4}
\usepackage[dvipdfm]{graphicx}
\usepackage{bm}

\begin{document}

\title{Edge Effect on Electronic Transport Properties of Graphene
 Nanoribbons and Presence of Perfectly Conducting Channel}
\author{Katsunori Wakabayashi$^{1,2}$}
\author{Yositake Takane$^1$}
\author{Masayuki Yamamoto$^1$}
\author{Manfred Sigrist$^{3}$}
\affiliation{$^1$Department of Quantum Matter, AdSM, Hiroshima University,
Higashi-Hiroshima 739-8530, Japan}
\affiliation{$^2$PRESTO, Japan Science and Technology Agency (JST),
Kawaguchi, Saitama 332-0012, Japan}
\affiliation{$^3$Theoretische Physik, ETH-H\"onggerberg, Z\"urich CH-8093,
Switzerland}

\vspace{30mm}
\noindent
$^\ast$ This paper is accepted for publication in CARBON(Elsevier).
\\
$^{\ast\ast}$
To whom all correspondencec should be addressed\\
Tel: 81-82-424-7654\\
Fax: 81-82-424-7000\\
E-mail: kwaka@hiroshima-u.ac.jp

\begin{abstract}
Numerical calculations have been performed to elucidate unconventional
electronic transport properties in disordered nanographene ribbons with
zigzag edges (zigzag ribbons). 
The energy band structure of zigzag ribbons has two valleys that are
well separated in momentum space, 
related to the two Dirac points of the graphene spectrum. 
The partial flat bands due to edge states make the imbalance 
between left- and right-going modes in each valley, {\it i.e.}
appearance of a single chiral mode.
This feature gives rise to a perfectly conducting channel in the
disordered system, i.e. the 
average of conductance $\langle g\rangle$ converges exponentially to 1
conductance quantum per 
spin with increasing system length, 
provided impurity scattering does not connect the two valleys, 
as is the case for long-range impurity potentials. 
Ribbons with short-range impurity potentials, however, through
inter-valley scattering, 
display ordinary localization behavior. 
Symmetry considerations lead to the classification of disordered zigzag
ribbons into 
the unitary class for long-range impurities, 
and the orthogonal class for short-range impurities. 
The electronic states of graphene nanoribbons with general edge
structures are also discussed, 
and it is demonstrated that chiral channels due to the edge states are 
realized even in more general edge structures
except for armchair edges. 
\end{abstract}

\maketitle

\section{Introduction}
Graphene being for many decades only a domain to theoretical studies, has recently been
fabricated with ingenious methods and has initiated intensive and diverse
research on this system.\cite{novoselov}
The honeycomb crystal structure of single layer graphene
consists of two inequivalent sublattices and results in a unique 
band structure for the itinerant $\pi$-electrons near
the Fermi energy which behave as massless Dirac fermion.
The valence and conduction bands touch conically at two nonequivalent
Dirac points, called $\bm{K_+}$ and $\bm{K_-}$ point,
which form a time-reversed pair, {\it i.e.} opposite chirality.\cite{geim}
The chirality and a Berry phase of $\pi$ at the two Dirac points 
provide an environment for highly unconventional and fascinating
two-dimensional electronic 
properties,  
such as the half-integer quantum Hall effect,\cite{qhe1,qhe2} 
the absence of backward scattering,\cite{ando.nakanishi}
$\pi$-phase shift of the Shubnikov-de Haas oscillations.\cite{kopelvich} 

\begin{figure}[h]
\includegraphics[width=0.8\linewidth]{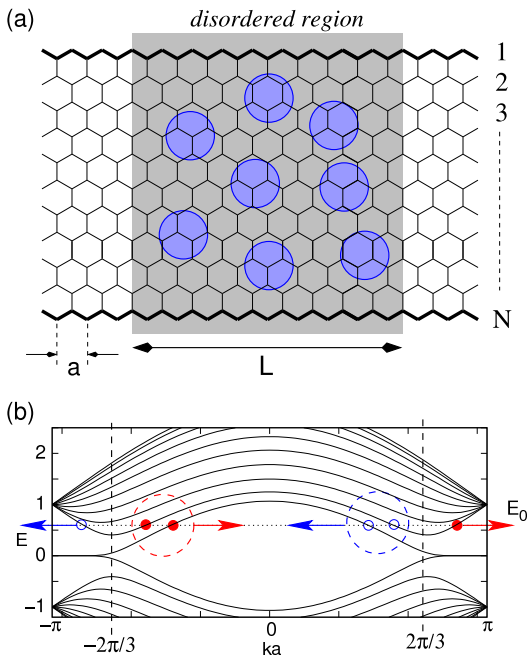}
\caption{(a) Structure of graphene zigzag ribbon. The disordered region with randomly
distributed impurities lies in the shaded region and has
the length $L$. The lattice constant is $a$ and the ribbon width $N$ is defined 
as the number of the zigzag chains.
Randomly distributed blue circles schematically represent the long
 ranged impurities. 
(b) Energy dispersion of zigzag ribbon with $N=10$. The valleys in the 
energy dispersion near $k=2\pi/3a$ ($k=-2\pi/3a$) originate from the
Dirac $\bm{K_+}$($\bm{K_-}$)-point of graphene.
The red-filled (blue-unfilled) circles denote the right (left)-moving
open channel at the energy $E_0$(dashed horizontal line). In the
 left(right) valley, the degeneracy  
between right and left moving channels is missing due to 
one excess right(left)-going mode. The time-reversal symmetry under the
intra-valley scattering is also broken.}
\label{fig:ribbon}
\end{figure}

The successive miniaturization of the graphene electronic devices inevitably
demands the clarification of edge effects on the electronic structures
and electronic transport properties of nanometer-sized graphenes. 
The presence of edges in graphene has strong implications
for the low-energy spectrum 
of the $\pi$-electrons.\cite{peculiar,prb.1999} 
There are two basic shapes of edges, {\it armchair} and {\it zigzag} which
determine the properties of graphene ribbons. 
It was shown that ribbons with zigzag edges (zigzag ribbon) possess 
localized edge states with energies close to
the Fermi level.\cite{peculiar,prb.1999,Phd}
These edge states correspond to the non-bonding configurations
 as can be seen by examining the analytic solution for
semi-infinite graphite with a zigzag edge for which the wave functions of the edge states reside on
one sublattice only.\cite{peculiar}
In contrast, edge states are completely absent for ribbons with armchair edges. 
Recent experiments support the evidence of edge localized states.\cite{enoki,fukuyama}
Also, graphene nanoribbons can experimentally be produced by using
lithography techniques.\cite{Kim}

The electronic transport through zigzag ribbons shows 
a number of intriguing phenomena 
such as zero-conductance Fano resonances,\cite{prl,prb}
vacancy configuration dependent transport,\cite{vacancy}
valley filtering,\cite{rycerz}, half-metallic conduction\cite{son} and
spin Hall effect.\cite{kane}
It is also expected that 
the edge states play an important role for the magnetic properties
in nanometer-sized graphite systems,
because of their relatively large contribution to the density of
states at the Fermi
energy.\cite{peculiar,prb.1999,kusakabe,sokada,harigaya,takai,palacios} 

Since the graphene nanoribbons can be viewed as a new class of 
the quantum wires, one might expect that random impurities 
inevitably cause the Anderson localization, i.e.
conductance decays exponentially with increasing system length $L$
and eventually vanishes in the limit of $L \to \infty$.
However, it was shown that carbon nanotubes with long-ranged impurities
possess a perfectly conducting channel.\cite{suzuura} 
Also, recently, the present authors reported that
nanographene ribbon with zigzag edges possess one perfectly
conducting channel if the impurity potentials are long-ranged, 
induced by electronic states which
originate from the edge states.\cite{prl2007}
Recent studies show
that perfectly conducting channels can be stabilized in two
standard universality classes.
One is the symplectic universality class with an odd number of conducting
channels,\cite{suzuura,takane1,takane2}
and the other is the unitary universality class with the imbalance between
the numbers of conducting channels in two propagating
directions.~\cite{prl2007,ohtsuki,takane4}
The symplectic class consists of systems having time-reversal symmetry
without spin-rotation invariance, while the unitary class is characterized
by the absence of time-reversal symmetry.~\cite{beenakker.rmt}

\begin{figure}[t]
\includegraphics[width=0.7\linewidth]{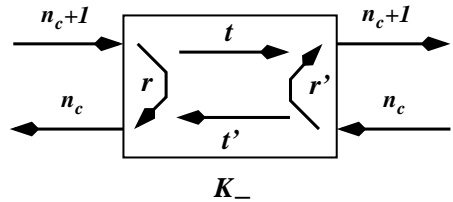}
\caption{Schematic figure of the scattering geometry at $\bm{K_-}$ point
in zigzag ribbons, where a single excess right-going channel
 exists. Here $n_c = 0,1,2,\cdots$.
On the contrary, the $\bm{K_+}$ has a single excess left-going mode.
}
\label{fig:rmat}
\end{figure}

In this paper, we study the disorder effects on the electronic transport
properties of graphene zigzag ribbons. The edge states play an important
role here, since they appear as special modes with partially flat bands
and lead under certain conditions  
to chiral modes separately in the two valleys. There is one such mode of opposite orientation 
in each of the two valleys of propagating modes,
which are well separated in $k$-space. The key result of this study is that for disorder
without inter-valley scattering a single perfectly conducting channel emerges
introduced by the presence of these chiral modes. This effect disappears
as soon as inter-valley scattering 
is possible. Therefore the behavior is determined by the range of the impurity potentials. 
As a function of the impurity potential range a crossover from the orthogonal to the unitary
universality class occurs which is connected with the presence or absence of 
time reversal symmetry (TRS).  
\begin{figure}
\includegraphics[width=0.6\linewidth]{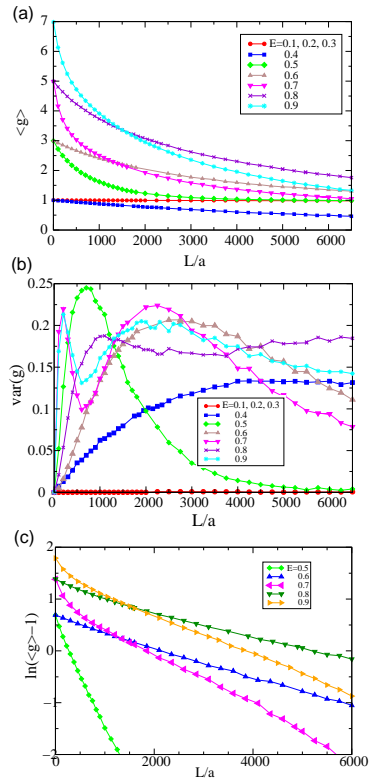}
\caption{$L$-dependence of (a) the
average, $ \langle g \rangle $, (b) the corresponding variance,
 $var(g)$, and (c) corresponding log-plot $\ln\langle g-1\rangle$ of
 dimensionless conductance  
for zigzag ribbon
with $N=10$,  $d/a=1.5$ (no inter-valley scattering),  $u_0=1.0$, and
$n_{imp.}=0.1$. More than 9000 samples with different impurity configuration
are included in the ensemble average.
}
\label{fig:aveg}
\end{figure}

We organize this paper as follows. In Sec. 2, we briefly introduce the
electronic states of graphene nanoribbons with zigzag edges.
In Sec. 3, some peculiar features in the scattering matrix formulation
on the graphene nanoribbons are explained, where the non-square form
of the reflection matrix and the implications of the perfectly conducting
channel in nanoribbons are presented. Also, the results of numerical
calculation are shown. 
In Sec. 4, we discuss the electronic states of graphene nanoribbons
with general edge structures, and demonstrate that chiral
channels due to the edge states is realized even in the general edge
structures except for the armchair edge.
In Sec. 5, the universality class of the nanographene ribbons is discussed.
The conclusion is presented in Sec. 6.

\section{Electronic states}
We describe the electronic states of
nanographites by the tight-binding model
\begin{eqnarray}
H = \sum_{i,j} \gamma_{i,j}|i\rangle\langle j| 
  + \sum_i V_i |i\rangle\langle i|, 
\label{eq:hamiltonian}
\end{eqnarray}
where $\gamma_{i,j}=-1$ if $i$ and $j$ are nearest neighbors, and 0 otherwise. 
$|i\rangle$ represents the state of the $p_z$-orbital 
on site $i$ neglecting the spin degrees of freedom. In the following we will
also apply magnetic fields perpendicular to the 
graphite plane which are incorporated via the Peierls phase: 
\begin{equation}
\gamma_{i,j}\rightarrow\gamma_{i,j}\exp\left[
i2\pi\frac{e}{ch}\int_i^jd\bm{l\cdot A}\right], 
\end{equation}
where $\bm{A}$
is the vector potential. 
The second term in Eq. (\ref{eq:hamiltonian}) represents the 
impurity potential, 
$V_i=V(\bm{r}_i)$ is the 
impurity potential at a position $\bm{r}_i$.

In Fig.\ref{fig:ribbon}(a), the graphite ribbon with zigzag edges (zigzag ribbons ) is shown. 
We assume edge sites are terminated by H-atoms.
The ribbon width $N$ is defined by the number of 
zigzag lines. The length of disordered region is defined as $L$.
Fig. \ref{fig:ribbon}(b) depicts
the energy band structure of zigzag ribbon for $N=10$. 
The zigzag ribbons are metallic for arbitrary
ribbon width. The most remarkable feature is the presence of
a partly flat band at the Fermi level, where the electrons
are strongly localized near the zigzag edge.
Each edge state has a non-vanishing amplitude
only on one of the two sublattices, having, thus, non-bonding character.
However, in a zigzag ribbon of finite width, two edge states coming
from both sides, have finite overlap. Because they are located on
different sublattices, they mix into a bonding and anti-bonding
configuration. In this way the partly flat bands acquire a
dispersion and become conductive except at exactly $ E=0 $.
Note that the overlap is increasing as $k$
deviates from $\pm \pi/a $, 
because the
penetration depth of the edge states 
increases and diverges at $ k = \pm 2\pi/3a $,
where $ a $ is the lattice constant.

We briefly discuss here the relation between valleys in the zigzag
ribbons and two-dimensional graphene. 
The electronic states near the Dirac point can be
described by the $\bm{k\cdot p}$ Hamiltonian 
\begin{equation}
H_{\bm{k\cdot p}}= \tilde{\gamma}
\left[
\hat{k}_x(\sigma^x\otimes \tau^0 )
+\hat{k}_y(\sigma^y\otimes\tau^z)
\right]
\end{equation}
acting on the 4-component pseudo-spinor Bloch functions
$\Phi(\bm{r})=\left[
\phi_{\bm{K_+}A},
\phi_{\bm{K_+}B},
\phi_{\bm{K_-}A},
\phi_{\bm{K_-}B}
\right]$, which characterize the wave functions on the two crystalline
sublattices (A and B) for the two Dirac points (valleys) $ \bm{K}_{\pm} $.
Here, $\tilde{\gamma}$ is the band parameter, $\hat{k}_x$($\hat{k}_y$)
are wavenumber operators, and $\tau^0 $ is the $2 \times 2 $ identity matrix.
Pauli matrices $\sigma^{x,y,z}$ act on the sublattice space ($A$, $B$),
while $\tau^{x,y,z}$ on the valley space ($\bm{K}_{\pm}$).
Since the outermost sites along $1^{st}$ ($N^{th}$) zigzag chain
are B(A)-sublattice, an imbalance between two sublattices occurs
at the zigzag edges leading to 
the boundary conditions
\begin{equation}
\phi_{\bm{K_\pm}A}(\bm{r}_{[0]})=0, \quad \phi_{\bm{K_\pm}B}(\bm{r}_{[N+1]})=0,
\end{equation}
where $\bm{r}_{[i]}$ stands for
the coordinate at $i^{th}$ zigzag chain.
It can be shown that
the valley near $k=3\pi/2a$ in Fig.1(b) originates
from the $\bm{K_+}$-point, the other valley at $k=-3\pi/2a$
from $\bm{K_-}$-point.\cite{Phd,note.bc,luis}
\begin{figure}
\includegraphics[width=0.6\linewidth]{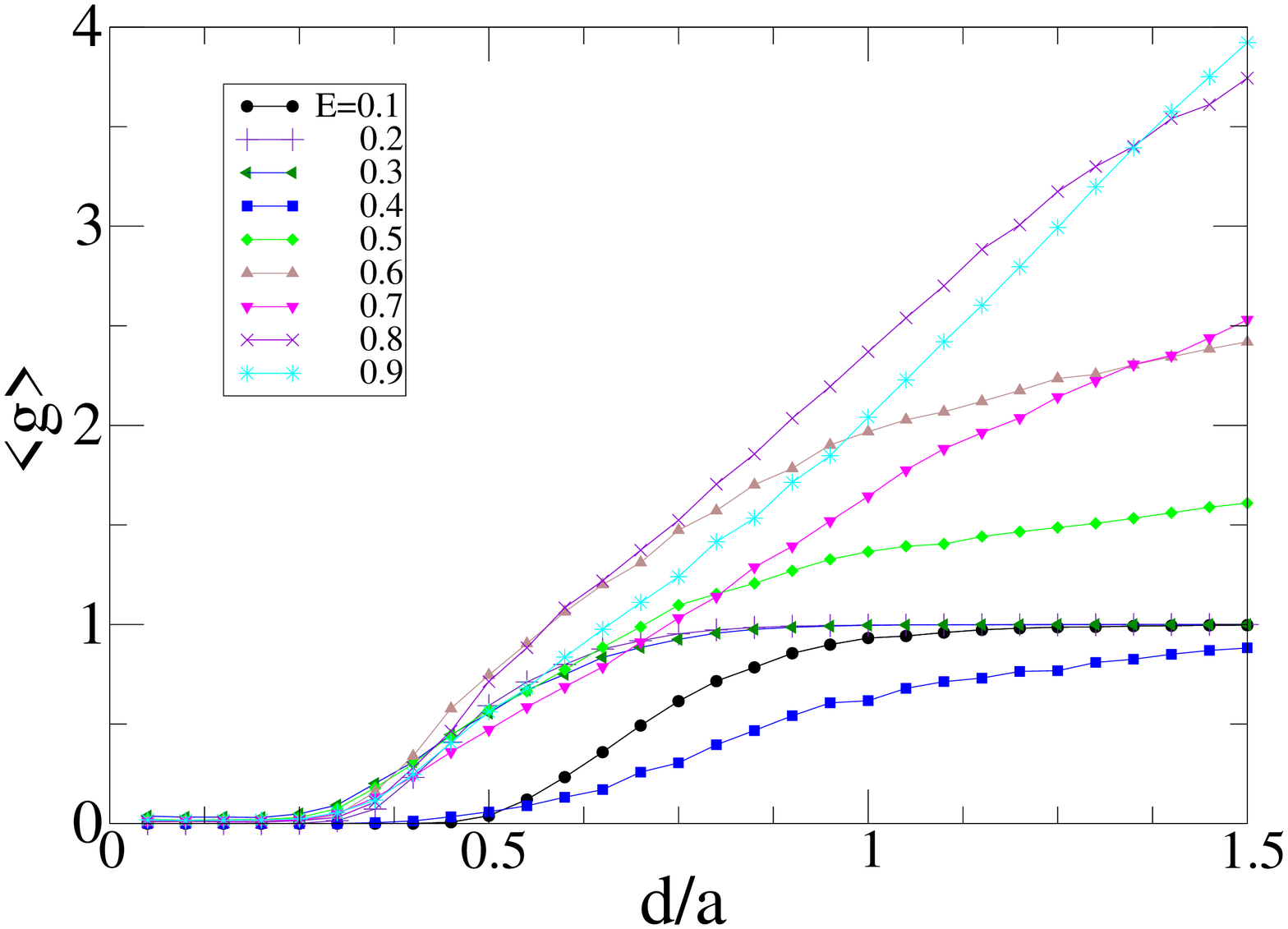}
\caption{Impurity range, $d/a$, dependence of the averaged dimensionless
conductance, $\langle g\rangle$, 
for zigzag ribbon with $N=10$, $u_0=1.0$, $L/a=5000$ and
$n_{imp.}=0.1$. More than 1000 samples with different impurity configuration
are included in the ensemble average.
}
\label{fig:dadep}
\end{figure}

Since the momentum difference between two
valleys is rather large, $\Delta k = k_+ - k_- =4\pi/3a$, 
only short-range impurities (SRI) with a range smaller than
the lattice constant causes {\it inter-valley scattering}. Long-range impurities (LRI), in contrast,
restrict the scattering processes to {\it intra-valley scattering}.
\cite{ando.nakanishi}

\begin{figure}
\includegraphics[width=0.9\linewidth]{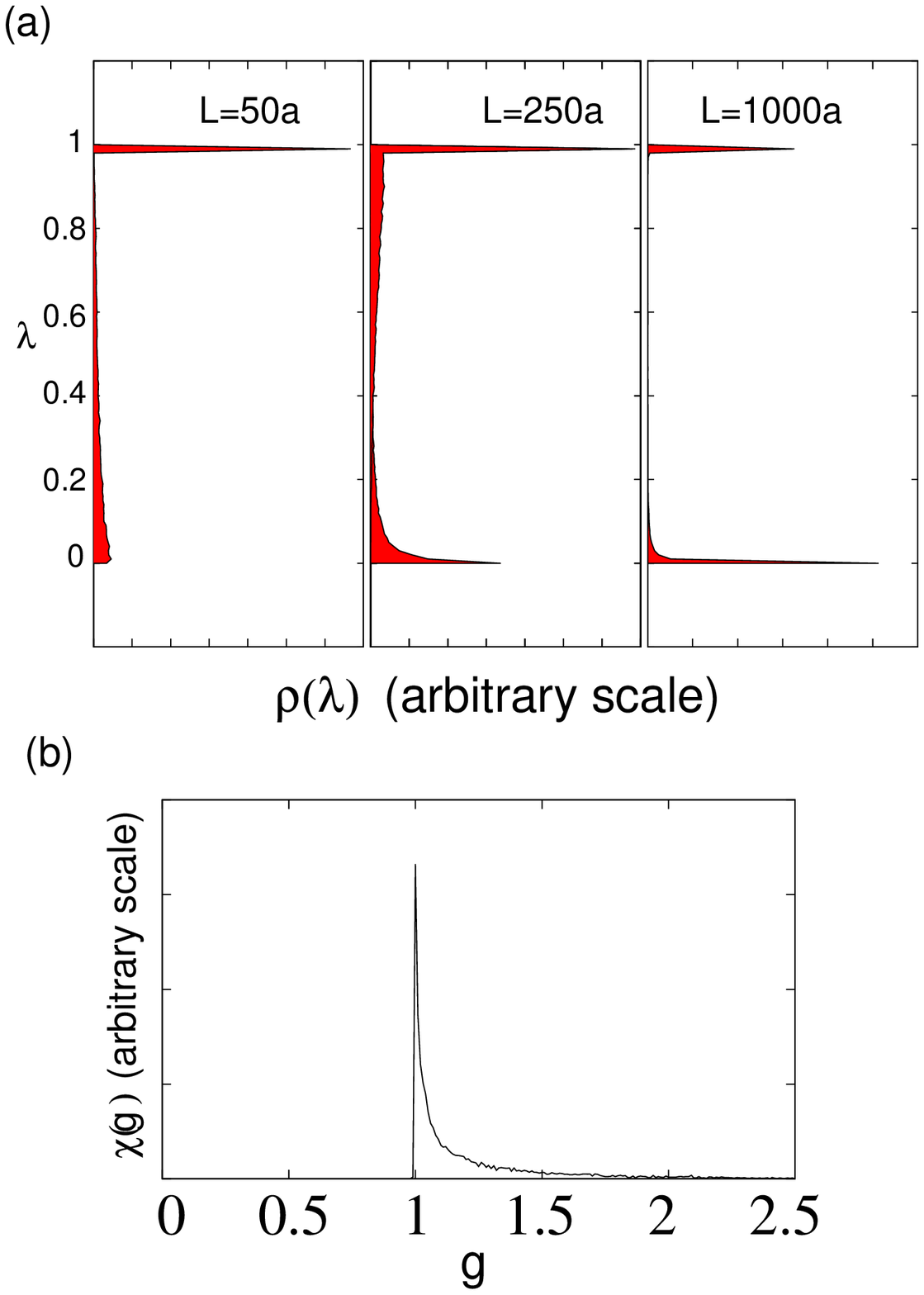}
\caption{(a) Distribution of the transmission eigenvalues $\lambda$: $\rho(\lambda)$,
at $E=0.5$ for $L/a=50,250,1000$, with $d/a=2.0$. $E=0.5 $ leads to 3 incident channels.
12,000 samples with different impurity configurations are included in the distribution.
(b) Distribution of the dimensionless conductance $g$: $\chi(g)$, at $L/a=1000$ 
 for the same parameter set. } 
\label{fig:eigen5}
\end{figure}

\begin{figure*}
\includegraphics[width=0.9\linewidth]{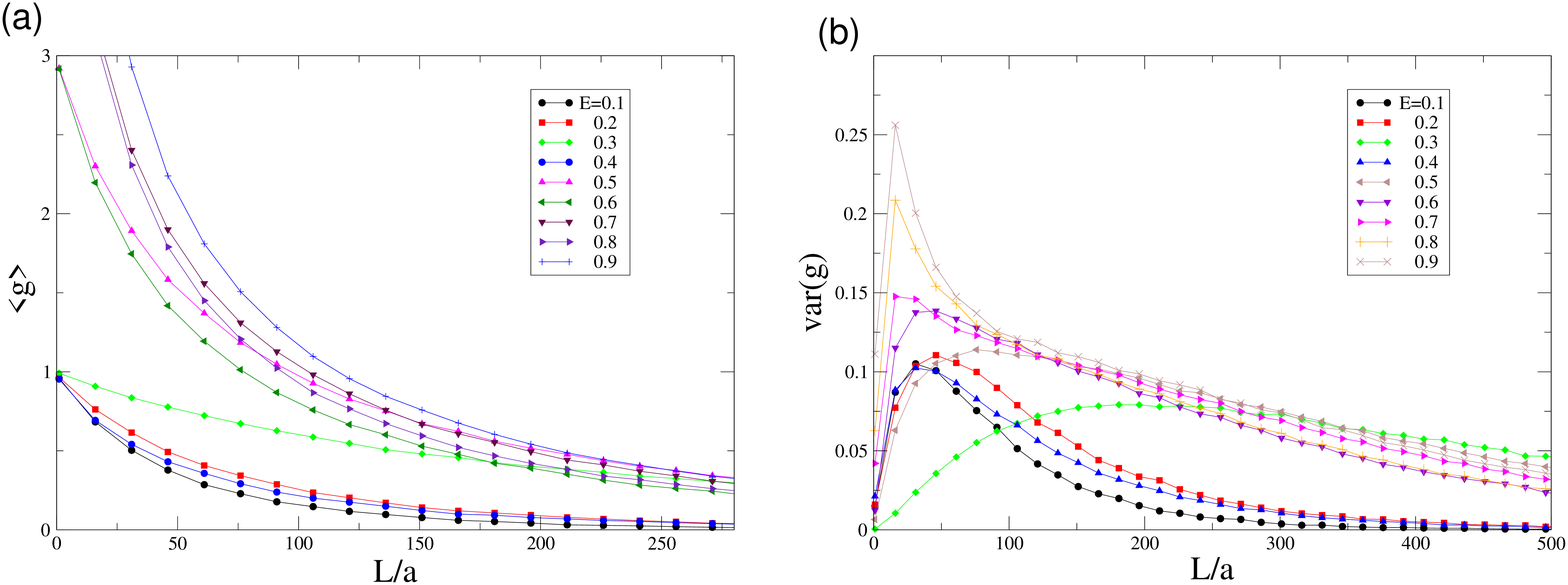}
\caption{$L$-dependence of (a) the
averaged dimensionless conductance and (b) the variance of the
 conductance for zigzag ribbons with 
$N=10$, short-ranged impurity potential ($d/a=0.05$, inter-valley scattering), 
$u_0=1.0$, and $n_{imp.}=0.1$. 20000 samples with different impurity configurations
are in the ensemble average.}
\label{fig:aveg_short}
\end{figure*}

In the graphene systems, pairs of time-reversed states
are formed across the two valleys (Dirac points).
In the absence of inter-valley scattering for LRI,
this ordinary TRS becomes irrelevant,
while the pseudo time-reversal symmetry with respect to the operator
$\mathcal{T} = - i (\sigma_{y} \otimes \tau_{0}) C$ ($C$: complex conjugation)
appears, where the A-B sublattices act as pseudospin.
This corresponds to the time-reversal operation restricted to each valley.
The boundary conditions which treat the two sublattices asymmetrically
leading to edge states give rise to a single special mode in each valley.
Considering now one of the two valleys separately,
say the one around $k = k_{+}$, we see that the pseudo TRS is violated
in the sense that we find one more left-moving than right-moving mode.
Thus, as long as disorder promotes only intra-valley scattering,
the system has no time-reversal symmetry.
On the other hand, if disorder yields inter-valley scattering,
the pseudo TRS disappears but the ordinary TRS is relevant making
a complete set of pairs of time-reversed modes across the two valleys.
Thus we expect to see qualitative differences in the properties
if the range of the impurity potentials is changed.
\begin{figure}
\includegraphics[width=0.8\linewidth]{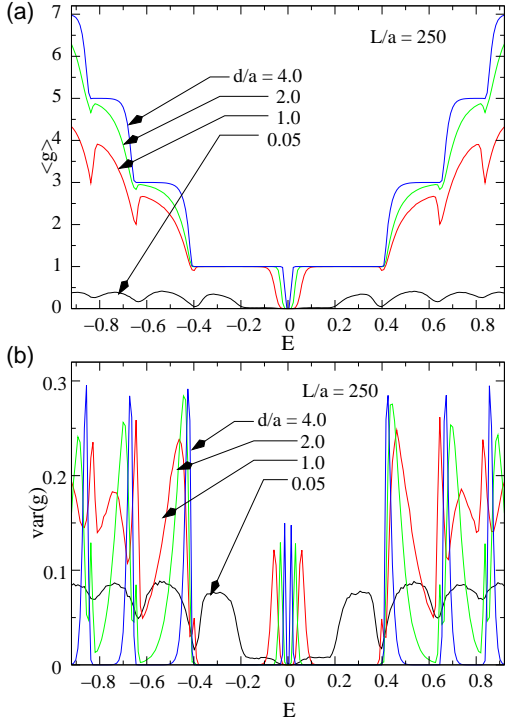}
\caption{
Fermi energy dependence of (a) the
averaged dimensionless conductance and (b) the variance of the
conductance for zigzag ribbons with 
$N=10$ for various impurity potential ranges at $L/a=250$.
10000 samples with different impurity configurations
are in the ensemble average.
Here, $u_0=1.0$ and $u_{imp.}=0.1$.
}
\label{fig:fermidep}
\end{figure}
\begin{figure}
\includegraphics[width=0.8\linewidth]{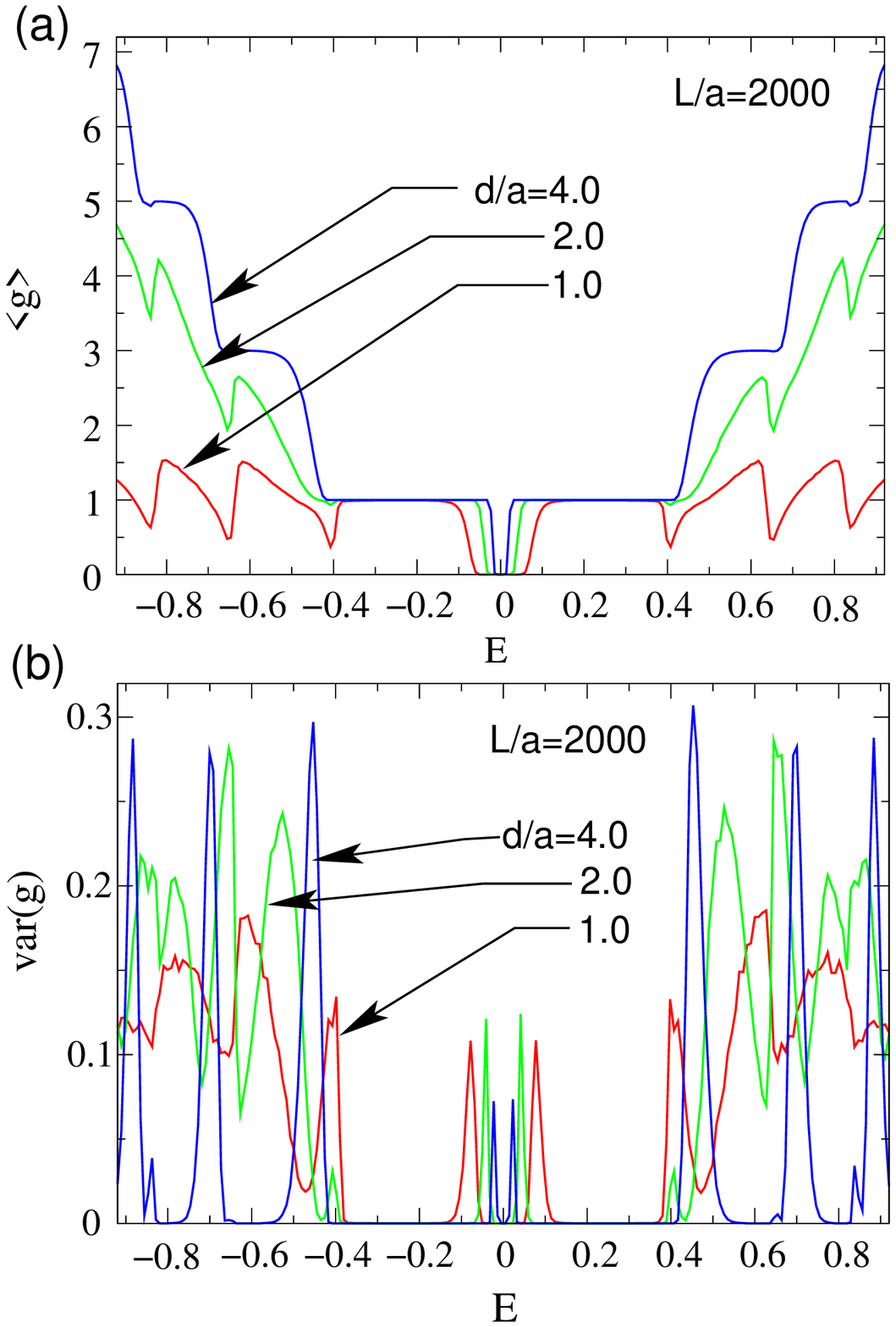}
\caption{
Fermi energy dependence of (a) the
averaged dimensionless conductance and (b) the variance of the
conductance for zigzag ribbons with 
$N=10$ for various impurity potential ranges at $L/a=2000$.
3200 samples with different impurity configurations
are in the ensemble average. In this regime the averaged conductance
for short-ranged impurities($d/a\ll 1$) is zero.
Here, $u_0=1.0$ and $u_{imp.}=0.1$.
}
\label{fig:fermidep4000}
\end{figure}

\section{Electronic transport properties}
\subsection{One-way excess channel system}
We numerically discuss the electronic transport properties of
the disordered nanographene ribbons. 
In general, electron scattering in a quantum wire is described by the
scattering matrix.\cite{beenakker.rmt}
Through the scattering matrix $\bm{S}$, the
amplitudes of the scattered waves $\bm{O}$ are
related to the incident waves $\bm{I}$,
\begin{equation}
\left(
\begin{array}{c}
\bm{O_L} \\
\bm{O_R}
\end{array}
\right)
=
\bm{S}
\left(
\begin{array}{c}
\bm{I_L} \\
\bm{I_R}
\end{array}
\right)
=\left(
\begin{array}{cc}
\bm{r} & \bm{t^\prime} \\
\bm{t} & \bm{r^\prime}
\end{array}
\right)
\left(
\begin{array}{c}
\bm{I_L} \\
\bm{I_R}
\end{array}
\right).
\end{equation}
Here, $\bm{r}$ and $\bm{r^\prime}$ are reflection matrices, 
$\bm{t}$ and $\bm{t^\prime}$ are transmission matrices,
$L$ and $R$ denote the left and right lead lines.
The Landauer-B\"uttiker formula\cite{mclbf} relates the
scattering matrix to the conductance of the sample. 
The electrical conductance is calculated using 
the Landauer-B\"uttiker formula,
\begin{equation}
G(E)
= \frac{e^2}{\pi\hbar}{\rm Tr}(\bm{t}\bm{t}^\dagger)
= \frac{e^2}{\pi\hbar} g(E).
\end{equation}
Here the transmission matrix $\bm{t}(E)$ can be calculated by means of
the recursive Green function method.\cite{prl,prb,green2}
For simplicity, throughout this paper, we evaluate electronic
conductance in the unit of quantum conductance ($e^2/\pi\hbar$), 
{\it i.e.} dimensionless conductance $g(E)$. 
We would like to mention that recently the edge disorder effect
on the electronic transport properties of graphene nanoribbons
was studied using similar approach.\cite{li,louis,mucciolo}

In the clean limit, the conductance of the zigzag ribbon can be given
simply by the number of the conducting channel.
As can be seen in Fig. \ref{fig:ribbon}(b), there is always one excess
left-going channel in the right valley ($\bm{K_+}$) within the
energy window of $|E|\le 1$. Analogously, there is one excess
right-going channel in the left valley ($\bm{K_-}$) within the same
energy window. Although the number of right-going and left-going channels
are balanced as a whole system, however if we focus on one of two valleys,
there is always one excess channel in one direction, {\it i.e.} chiral mode.

Now let us inject electrons from left to right-side
through the sample. When the chemical potential is changed from $E=0$, the
quantization rule of the dimensionless conductance ($g_{\bm{K_+}}$) in
the valley of $\bm{K_+}$ is given as
\begin{equation}
g_{\bm{K_+}} = n,
\end{equation}
where $n=0,1,2,\cdots$.
The quantization rule in the $\bm{K_-}$-valley is
\begin{equation}
g_{\bm{K_+}} = n+1.
\end{equation}
Thus, conductance quantization of the zigzag ribbon in the clean
limit near $E=0$ has the following odd-number quantization, i.e. 
\begin{equation}
g=g_{\bm{K_+}}+g_{\bm{K_-}}= 2n+1.
\end{equation}

Since we have an excess mode in each valley, the scattering matrix
has some peculiar features which can be seen when we
explicitly write the valley dependence 
in the scattering matrix.
By denoting the contribution of the right valley ($\bm{K_+}$) as $+$,
and of the left valley  ($\bm{K_-}$) as $-$, the scattering matrix
can be rewritten as 
\begin{equation}
\left(
\begin{array}{c}
\bm{O^+_L} \\
\bm{O^-_L} \\
\bm{O^+_R} \\
\bm{O^-_R}
\end{array}
\right)
=\left(
\begin{array}{cc}
\bm{r} & \bm{t^\prime} \\
\bm{t} & \bm{r^\prime}
\end{array}
\right)
\left(
\begin{array}{c}
\bm{I^+_L} \\
\bm{I^-_L} \\
\bm{I^+_R} \\
\bm{I^-_R}
\end{array}
\right).
\end{equation}
Here we should note that the dimension of each column vector is
not identical. Let us denote the number of the right-going channel 
in the valley $\bm{K_+}$ or the left-going channel in the
valley $\bm{K_-}$ as $n_c$. For example,
$n_c = 1$ at $E=E_0$ in the Fig.\ref{fig:ribbon}(b).
Thus the dimension of the column vectors is given as follows:
\begin{equation}
\left\{
\begin{array}{ll}
dim(\bm{I^+_L}) = n_c,     & dim(\bm{I^+_R}) = n_c + 1, \\ 
dim(\bm{I^-_L}) = n_c + 1, & dim(\bm{I^-_R}) = n_c,     \\ 
\end{array}
\right.
\end{equation}
and 
\begin{equation}
\left\{
\begin{array}{ll}
dim(\bm{O^+_L}) = n_c + 1, & dim(\bm{O^+_R}) = n_c,    \\ 
dim(\bm{O^-_L}) = n_c,     & dim(\bm{O^-_R}) = n_c + 1. \\ 
\end{array}
\right.
\end{equation}
Subsequently, the reflection and transmission matrices have the following
matrix structures, 
\begin{eqnarray}
\bm{r}= \bordermatrix{
      & n_c       & n_c+1         \cr
n_c+1 & \bm{r_{++}} & \bm{r_{+-}} \cr
n_c   & \bm{r_{-+}} & \bm{r_{--}}
},
\end{eqnarray}
\begin{eqnarray}
\bm{t}= \bordermatrix{
      & n_c       & n_c+1         \cr
n_c   & \bm{t_{++}} & \bm{t_{+-}} \cr
n_c+1 & \bm{t_{-+}} & \bm{t_{--}}
},
\end{eqnarray}
\begin{eqnarray}
\bm{r^\prime}= \bordermatrix{
      & n_c+1      & n_c         \cr
n_c & \bm{r^\prime_{++}} & \bm{r^\prime_{+-}} \cr
n_c+1   & \bm{r^\prime_{-+}} & \bm{r^\prime_{--}}
},
\end{eqnarray}
\begin{eqnarray}
\bm{t^\prime}= \bordermatrix{
      & n_c+1       & n_c            \cr
n_c+1 & \bm{t^\prime_{++}} & \bm{t^\prime_{+-}} \cr
n_c   & \bm{t^\prime_{-+}} & \bm{t^\prime_{--}}
}.
\end{eqnarray}
The reflection matrices
become non-square when the intervalley scattering is suppressed, {\it i.e.}
the off-diagonal submatrices are zero.

When the electrons are injected from the
left lead of the sample and the intervalley scattering is suppressed, 
a system with an excess channel is realised in the $\bm{K_-}$-valley.
The scattering geometry is schematically drawn in Fig.\ref{fig:rmat}.
Thus, for single valley transport,
the $\bm{r_{--}}$ and $\bm{r^\prime_{--}}$ are $n_c\times (n_c+1)$ and $(n_c+1)\times n_c$
matrices, respectively, and $\bm{t_{--}}$ and $\bm{t^\prime_{--}}$ are
$(n_c+1)\times (n_c + 1)$ and $n_c \times n_c$ matrices, respectively.
Noting the dimensions of $\bm{r_{--}}$ and $\bm{r^\prime_{--}}$, we find that
$\bm{r_{--}}^\dagger \bm{r_{--}}$ and $\bm{r^\prime_{--}}
{\bm{{r^\prime}^\dagger_{--}}}$ have a single 
zero eigenvalue. Combining this with the flux conservation relation
($\bm{S}^\dagger \bm{S} = \bm{S}\bm{S}^\dagger = \bm{1}$), we arrive at
the conclusion that 
$\bm{t_{--}t^\dagger_{--}}$ has an eigenvalue equal to unity.
Note that 
${\bm {t^\prime}_{--}} {\bm {t^\prime}}^\dagger_{--} $
does not
have such an anomalous eigenvalue. 
This indicates the presence of a perfectly conducting channel (PCC)
only in the right-moving channels. 
If the set of eigenvalues for 
${\bm {t^\prime}_{--}} {\bm {t^\prime}}^\dagger_{--}$ is expressed as
$\{T_1, T_2, \cdots, T_{n_c}\}$, that for
${\bm{t_{--}}{\bm t}^\dagger_{\bm --}}$ is
expressed as $\{T_1, T_2, \cdots, T_{n_c}, 1\}$, i.e. a PCC.
Thus, the dimensionless conductance g for the right-moving channels is
given as 
\begin{equation}
g_{\bm{K_-}}=\sum_{i=1}^{n_c+1}T_i = 1 + \sum_{i=1}^{n_c}T_i,
\end{equation}
while that for the left-moving channels is 
\begin{equation}
g^\prime_{\bm{K_-}} = \sum_{i=1}^{n_c}T_i. 
\end{equation}
We see  that $g_{\bm{K_-}}=g^\prime_{\bm{K_-}} +1$.
Since the overall TRS of the system guarantees
the following relation:
\begin{equation}
\begin{array}{c}
g^\prime_{\bm{K_+}} = g_{\bm{K_-}},\\
g^\prime_{\bm{K_-}} = g_{\bm{K_+}},
\end{array}
\end{equation}
the conductance 
$g=g_{\bm{K_+}}+g_{\bm{K_-}}$ (right-moving) and 
$g^\prime=g^\prime_{\bm{K_+}}+g^\prime_{\bm{K_-}}$ (left-moving) are equivalent.
If  the probability distribution of $\{T_i\}$ is obtained as a function
$L$, we can describe the statistical properties of $g$ as well as
$g^\prime$.
The evolution of the distribution function with increasing $L$ is
described by the DMPK(Dorokhov-Mello-Pereyra-Kumar) equation for 
transmission eigenvalues.\cite{takane4}

In the following, the presence of a perfectly conducting
channel in disordered nanographene ribbons will be demonstrated 
with the help of numerical calculation. 
Recently Hirose {\it et. al.} pointed out that the
Chalker-Coddington model which possesses non-square 
reflection matrices with unitary symmetry gives rise to a perfectly
conducting channel.\cite{ohtsuki}
However, systems with an excess channel in one direction has been believed difficult to realize. 
Therefore disordered nanographene ribbons with LRI  might constitute the first
realistic example. It is possible to extend the discussion 
to generic multiple-excess channel model, where the $m$-PCCs appear.\cite{takane4}
Such systems can be realized by stacking zigzag nanographene
ribbons.\cite{miyamoto}
The electronic transport due to PCC resembles to the electronic
transport due to a chiral mode in quantum 
Hall system.\cite{macdonald,ishizaka} 
However, it should be noted that the PCC due to edge states 
in zigzag ribbons occurs even without the magnetic field.

\subsection{model of impurity potential}
In our model  we assume that the impurities are randomly distributed with a density $n_{imp}$, and
the potential has a Gaussian form of a range $d$
\begin{equation}
V(\bm{r}_i) = \sum_{\bm{r_0}(random)}u
\exp\left(-\frac{|\bm{r}_i-\bm{r}_0|^2}{d^2}\right) 
\end{equation}
where the strength $u$ is uniformly distributed within the range $|u|\le u_{M}$.
Here $u_{M}$ satisfies the normalization condition:
\begin{equation}
u_{M}\sum_{{\bm r}_i}^{(full\ space)}
\exp\left(-{\bm{r}_i^2}{d^2}\right)/(\sqrt{3}/2)=u_0.
\end{equation}
Since the momentum difference between two
valleys is rather large, $\Delta k = k_+ - k_- =4\pi/3a$, 
only short-range impurities (SRI) with a range smaller than
the lattice constant causes {\it inter-valley scattering}. Long-range impurities (LRI), in contrast,
restrict the scattering 
processes to {\it intra-valley scattering}.\cite{ando.nakanishi}

\subsection{disordered nanographene ribbons}
We focus first on the case of LRI
using a potential with $ d/a=1.5 $ which is already sufficient to avoid inter-valley scattering. 
Fig.\ref{fig:aveg} shows (a) the averaged dimensionless conductance 
and (b) the corresponding variance as a
function of $ L $ for different incident energies, averaging over an
ensemble of 40000 samples with 
different impurity configurations for ribbons of the width $N=10$.
The variance, which describes the fluctuation of the conductance, is defined as
\begin{equation}
var(g) = \langle g^2\rangle - \langle g\rangle^2. 
\end{equation}
The potential strength and impurity density are chosen to be 
$u_0=1.0$ and $n_{imp.} = 0.1$, respectively. As a typical localization effect we observe
that $\langle g \rangle $ gradually decreases with growing length $ L $ (Fig.\ref{fig:aveg}).
Surprisingly, the ribbons remain highly conductive even at the 
length of $L=1500a$, i.e. more than $350nm$ in the real system. Actually, $ \langle g \rangle $ 
converges to $\langle g\rangle =1$, indicating the presence of a single {\it perfectly conducting} 
channel. It can be seen that $\langle g\rangle(L) $ has an exponential behavior as
\begin{equation}
\langle g\rangle -1 \sim \exp(-L/\xi)
\end{equation}
with $\xi$ as the localization length. 
In this paper, the localization length is evaluated, by identifying
$\exp\langle\ln \tilde{g}\rangle=\exp(-L/\xi)$. 
Here, $\tilde{g}=g-1$ ($\tilde{g}=g$)
for the system with (without) the perfectly conducting channel.

Interestingly, the variance for the LRI case shown in
Fig.\ref{fig:aveg}(b) has large values and slowly converges to zero
toward the long wire limit. Also, the variance for higher energy modes
has double humps structure in the short wire regime, which is also
indicating the unconventional behavior.
Such suppression of the fluctuation in the diffusive regime 
may be attributed to the level repulsion from the
transmission eigenvalue of the PCC.\cite{preprint}
Note the variance is zero for $E=0.1,0.2,0.3$,
indicating the existence of the perfectly conducting channel.
Such peculiar features cannot be seen for SRI case, see
Fig.\ref{fig:aveg_short}(b) for comparison.

Fig.\ref{fig:dadep} shows impurity range ($d/a$) dependence of the
averaged conductance for various Fermi energy. Clearly, it can be
seen that the PCC develops in zigzag ribbons 
if the range of impurity gets larger than the lattice constant. 
However, the development of PCC gets moderate 
if the incident energy lies at a value close to the change
between $ g = 2n-1 $ and $ g=2n+1 $  
for the ribbon without disorder. Since at least one of subbands 
at these energy points gives zero group velocity, it can be
considered that the intra-valley scattering becomes stronger. 
This feature is also for example visible in above
calculations as the deviation from the limit $ \langle g \rangle \to 1 $  
for $E =0.4 $ 
where the limiting value $ \langle g \rangle < 1 $ (Fig.\ref{fig:aveg}). 
This feature can be seen clearly in Fig.\ref{fig:fermidep} and
\ref{fig:fermidep4000}, which will be discussed later.

We performed a number of tests to confirm the presence of this perfectly
conducting channel.  
First of all, it exists up to $L=3000a$ for various ribbon widths up to
$N=40$ for the 
potential range ($d/a=1.5$). Moreover the perfectly conducting
channel remains for LRI with
$d/a=2.0,4.0,6.0,8.0$, and $u_0=1.0$,  $n_{imp.}=0.1$ and $ N=10 $. As
the effect is connected with the subtle feature of an excess mode in the
band structure, it is natural that the result
can only be valid for sufficiently weak potentials. For potential strengths comparable to the
energy scale of the band structure, e.g. the energy difference between the transverse modes, 
the result should be qualitatively altered.\cite{vacancy} 

As a further test we evaluate the distribution of the transmission eigenvalues
and dimensionless conductance for fixed wire length.
In Fig.\ref{fig:eigen5}(a), the distribution of the eigenvalues $\lambda$ of
the Hermite matrix, $\bm{tt}^\dagger$, is depicted for 
various wire lengths. With growing length $ L $ a progressive separation
of the transmission eigenvalues emerges with a strong peak close to 0
(localization) and at 1 (perfect conduction channel). 
The distribution of the conductance $ g $ (trace of the transmission matrix  
${\rm Tr}(\bm{tt}^\dagger)$), is depicted in Fig.\ref{fig:eigen5}(b)
for samples in the long-wire limit. 
Obviously, $ g $ only distributes above $g=1$ with a singularity at 1. 
\begin{figure}
\includegraphics[width=0.6\linewidth]{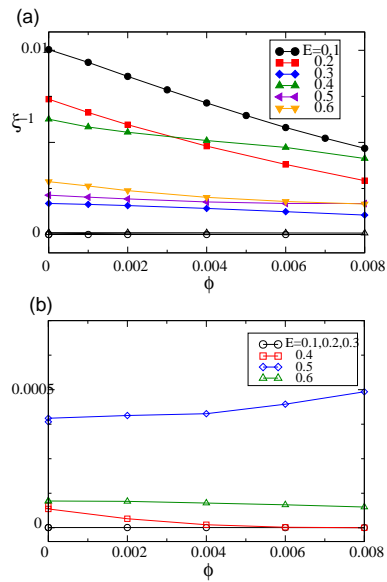}
\caption{(a) Magnetic field dependence of the inverse
 localization length $\xi^{-1}$ for various incident energies.
The filled symbols indicate data sets of systems with SRI being rather sensitive to
magnetic fields, and empty symbols denote data sets for LRI which are almost
insensitive to the field. The magnetic flux $ \phi$ passing through a single
hexagon ring is measured in units of the quantum flux ($\phi_0=ch/e$).
(b) Enlarged view of (a) of LRI data 
5000 samples with different configurations are included in the averages.}
\label{fig:xi}
\end{figure}

Turning to the case of SRI the inter-valley scattering becomes sizable enough
to ensure TRS, such that the perfect transport supported by the
effective chiral mode in a single valley ceases to exist. 
For a comparison, we show the ribbon length dependence of the averaged conductance in 
Fig.\ref{fig:aveg_short}. 
Since SRI causes the inter-valley scattering for any incident energy, 
the electrons tend to be localized and 
the averaged conductance decays exponentially, $\langle g\rangle\sim \exp(-L/\xi)$, without
developing a perfect conduction channel.

In Fig.\ref{fig:fermidep}, Fermi energy dependence of (a) the
averaged dimensionless conductance and (b) the variance of the
conductance for zigzag ribbons with 
$N=10$ for various impurity potential ranges $d/a=0.05, 1.0, 2.0, 4.0$
at $L/a=250$. 10000 samples with different impurity configurations
are in the ensemble average. With decreasing the range of the impurity
potential, the averaged conductance also gradually decreases and
the perfect conduction disappears for  short-ranged impurities of
$d/a=0.05$. As we have already briefly mentioned, the sharp dips
can be observed at energies where the number of conducting channels
changes. Also, the variances becomes relatively large at these energies.
The drop of the conductance close to E=0 indicates that the two valleys
are not well-separated in $k$-space, because the partial flat bands have finite
curvature for narrow graphene ribbons such as $N=10$. Therefore, if the
ribbon width widens, the perfectly conduction recovers even close to
the zero-energy.
In Fig.\ref{fig:fermidep4000}, similar figure for $L/a=2000$.

In order to demonstrate that the qualitative difference between the two regimes, LRI and SRI, is
indeed connected with TRS, we study the effect of magnetic field coupling to
the electrons through the Peierls phase. For the time reversal symmetric
situation resulting from SRI scattering the magnetic field removing TRS
should have a stronger effect than for the case of LRI where TRS is
broken already at the outset. We use the localization length $ \xi $ as
an indicator. In Fig.\ref{fig:xi}, the field  
dependence of the inverse localization length is shown for various
incident energies (filled symbols for SRI and empty symbols for
LRI). Indeed the localization length displays a stronger field
dependence than the LRI. Actually for LRI even a so-called
anti-localization behavior with increasing field is visible consistent
with recent reports on graphene.\cite{antilocalization,suzuura.prl,prb}
Note that for $ E < 0.4 $ only a single channel is involved in the 
conductance such that for LRI no localization occurs, i.e. $ \xi^{-1} = 0 $.

\section{General edge structures}

As we have seen, zigzag ribbons with long-ranged impurity potentials 
retain a single PCC. This PCC originates for the following two reasons: 
(i) The spectrum contains two valleys (two Dirac $\bm{K_\pm}$-points) which 
are well enough separated in
momentum space as to suppress intervalley scattering
due to the long-ranged impurities,
(ii) the spectrum in each valley is chiral by possessing a right- and left-moving
modes which differ by one in number, and so scattered electrons can 
avoid in one channel backscattering.
Is the zigzag ribbon the only nanographene ribbon showing this effect?
Here we would like to show that such conditions can be satisfied in 
the graphene nanoribbon with more general edge structures except for one
case, the armchair edge.
In the following we will first show that only the armchair edge cannot
produce the localized edge states. Moreover, the spectrum of the armchair
ribbon overlays the two Dirac $\bm{K}$-points at the single momentum  $k=0$
and displays therefore no separation into two valleys unlike the zigzag ribbon. 
We will then show that general edges can produce the necessary conditions to observe a
PCC in a disordered system.

\subsection{Continuum approach}

\begin{figure*}
\includegraphics[width=0.9\linewidth]{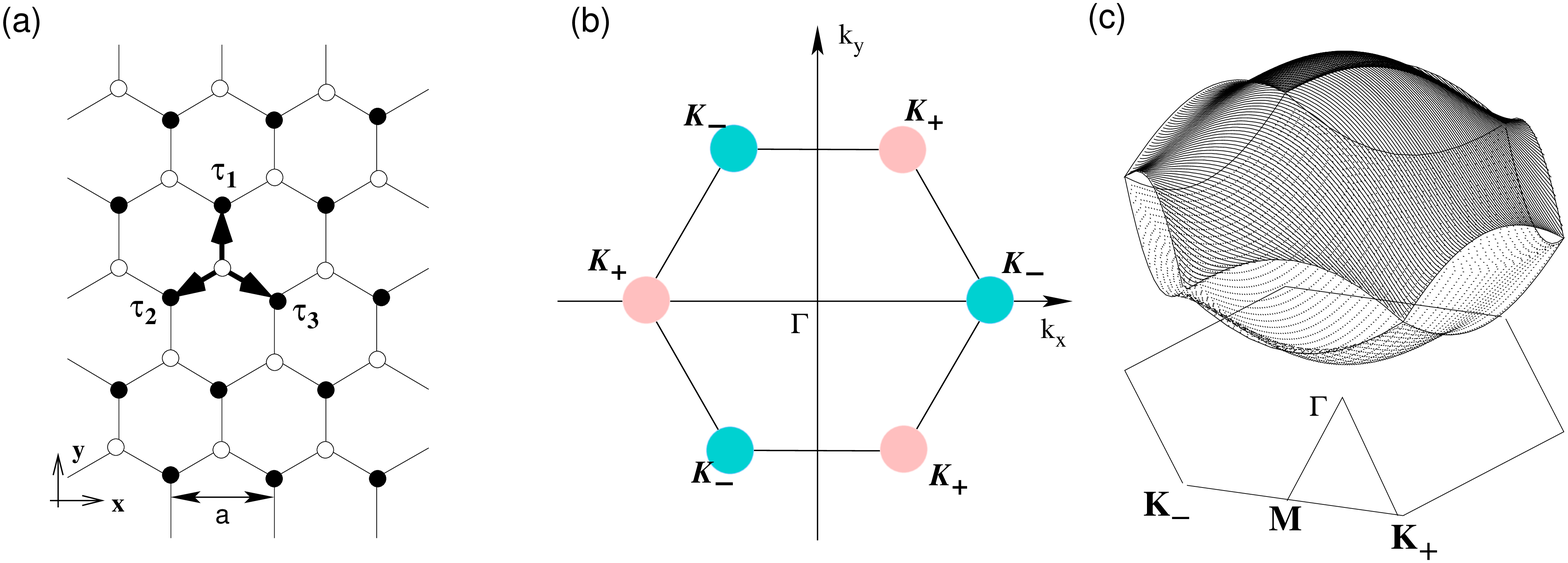}
\caption{(a) The lattice structure of graphene and the definition of
 $\tau_i$ ($i=1,2,3$) and coordinates.
Here 
$\bm{\tau}_1=a(   0, 1/ \sqrt{3})$,
$\bm{\tau}_2=a(-1/2,-1/2\sqrt{3})$, and
$\bm{\tau}_3=a( 1/2,-1/2\sqrt{3})$. 
(b) The corresponding 1st Brillouin Zone. 
$\bm{K}_{\pm}=\frac{2\pi}{a}( \pm \frac{1}{3},\frac{1}{\sqrt{3}})$,
 $\bm{\Gamma}=(0,0)$, and so on.
(c) The band energy structures for $\pi$ electrons.}
\label{fig:graphene}
\end{figure*}

We consider graphene at half-filling in order to explore their
zero-energy edge states. 
Here from the tight-binding model we derive the stationary Schr\"odinger
equation for graphene in momentum space,
\begin{equation}
\left[
\begin{array}{cc}
0 &\epsilon_{\bm k}^\ast \\
\epsilon_{\bm k} & 0 
\end{array}
\right] \hat{\Psi}_{\bm k}
=
E\hat{\Psi}_{\bm k}
\end{equation}
where $\epsilon_{\bm k}=-t\sum_i \exp(i\bm{k\cdot \tau_i})$ 
($\bm{\tau}_1=a(0,1/\sqrt{3})$,
$\bm{\tau}_2=a(-1/2,-1/2\sqrt{3})$, and
$\bm{\tau}_3=a( 1/2,-1/2\sqrt{3})$) and 
\begin{equation}
\hat{\Psi}_{\bm k} = \left( \begin{array}{c} \psi_A ({\bm k}) \\ \psi_{B} ( {\bm k}) \end{array}
\right)
\end{equation}
with $\psi_A(\bm{k})$ and $\psi_B(\bm{k})$ are the wavefunctions located
on the A- and B-sublattice, respectively.  
The spectrum contains the well-known linear Dirac spectrum
at two nonequivalent momentum $\bm{K_\pm}$ points,
$\bm{K}_{\pm}=\frac{2\pi}{a}( \pm \frac{1}{3},\frac{1}{\sqrt{3}})$.
In Fig.\ref{fig:graphene}, (a) the lattice structure of graphene and the
definition of $\bm{\tau}_i$($i=1,2,3$) and coordinates, (b)
corresponding 1st Brillouin Zone, and (c) $\pi$ band structures are shown.

In a first discussion concerning edge states we perform the following
transformation:
\begin{equation}
\hat{\Phi}_{\bm k} = \left( 
\begin{array}{c} 
\phi_p (\bm{k}) \\ 
\phi_h (\bm{k}) 
\end{array} 
\right) = 
\frac{1}{\sqrt{2}} 
\left(
\begin{array}{c} 
 \psi_A (\bm{k} ) + \psi_B(\bm{k})  \\
i\psi_B (\bm{k}) - i \psi_A(\bm{k})  
\end{array} 
\right)
\end{equation}
which yields the Schr\"odinger equation,
\begin{equation}
(\epsilon^+_{\bm k}\hat{\sigma}_z -i \epsilon^-_{\bm k}\hat{\sigma}_x)
\hat{\Phi}_{\bm k}=E\hat{\Phi}_{\bm k},
\end{equation}
where $\epsilon^\pm_{\bm k}= (\epsilon_{\bm k} \pm \epsilon_{\bm
k}^\ast)/2$. The structure of this equation is identical to that of a
BCS problem in particle-hole space $ (\psi_p , \psi_h) $, constituting a Bogolyubov-de Gennes 
equation.  The diagonal terms, $\epsilon^+_{\bm{k}}={\rm
Re}(\epsilon_{\bm{k}})$, formally correspond to the kinetic energy for
particles and holes, 
and the off-diagonal terms, $-i\epsilon^-_{\bm{k}}={\rm
Im}(\epsilon_{\bm{k}})$, can be considered as the pair-potential of a 
''superconductor''. 
\begin{figure}
\includegraphics[width=0.8\linewidth]{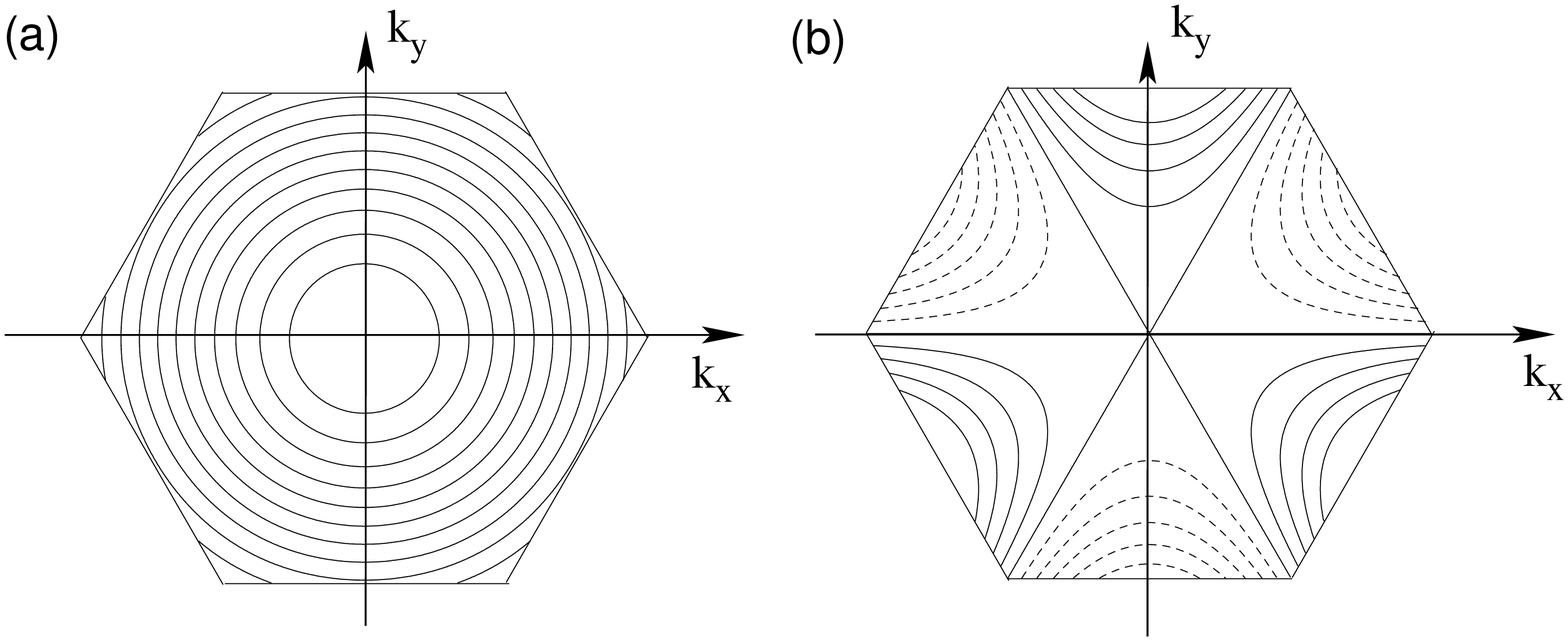}
\caption{The contour plot for (a) $\epsilon^+_{\bm{k}}$ (b)
 $\Delta_{\bm{k}}$. The dashed line means the negative values.}
\label{fig:delta.contour}
\end{figure}
\begin{figure}
\includegraphics[width=0.5\linewidth]{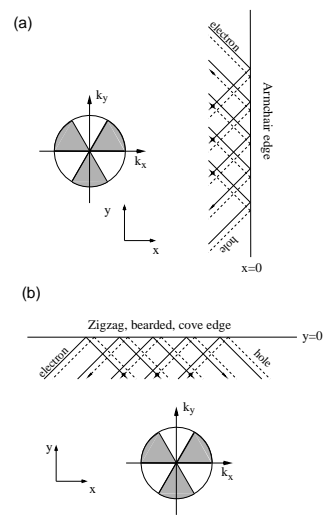}
\caption{Schematic view of the electron scattering in the surface region
 of graphene with the analogy to the Andreev reflection, where the
 electronic states of graphene were mapped to the picture of the $f$-wave
superconductors(see text). 
Circles with shading shows the wavenumber dependence of 
$\Delta_{\bm k}$, where shading means negative values.  
(a) The scattering trajectories at the armchair edge, which convert the
momentum as $k_x\rightarrow -k_x$, do not connect gap regions
of opposite sign.
However, (b) The scattering trajectories at the zigzag, bearded and cove
 edge, which convert the momentum as $k_y\rightarrow -k_y$, 
connect gap regions of opposite sign which lead to a phase shift $\pi$
and zero-energy bound states.
}
\label{fig:f-wave}
\end{figure}
Thus we rewrite the equation as
\begin{equation}
(\hat{\xi}_{\bm k}\hat{\sigma}_z + \Delta_{\bm k}\hat{\sigma}_x)
\hat{\Phi}_{\bm k}=E\hat{\Phi}_{\bm k} \; .
\end{equation}
In the continuum limit we can approximate
$\hat{\xi}_{\bm k} = - ({\hbar^2}/{2m})\nabla^2-\mu$, where $\mu$ is the chemical potential, and 
$\Delta_{\bm k}$ represents the pair-potential of
the superconductor,
$\Delta_{\bm k}=\left(\epsilon_{\bm k} - \epsilon_{\bm
k}^\ast \right)/2i={\rm Im}(\epsilon_{\bm{k}})$.
In Fig.\ref{fig:delta.contour}, 
the contour plot for (a) $\epsilon^+_{\bm{k}}$ (b) $\Delta_{\bm{k}}$
are shown. The pair potential shows line nodes along the momenta
$k_y = \pm \sqrt{3} k_x$ and $k_y=0$. Thus close to the Fermi energy 
we approximate the pair potential simply by its angular dependence
\begin{equation}
\Delta = |\Delta|\cos(3\theta),
\end{equation}
where $\theta$ is the angle of the Fermi vector relative to  
the positive $k_x$-axis. Interestingly, the ''pairing'' symmetry is 
odd-parity, an $f$-wave state.

From this properties we use now the general rules for the presence of 
zero-energy edge states.\cite{hu,kashiwaya} A related discussion can be found in  ref.\cite{ryu}. 
Considering a classical trajectory, the existence of a zero-energy bound state requires 
that the momentum incident to the edge and the momentum of specular scattered outgoing
trajectory lie on the Fermi surface on points which have a phase difference of $ \pi $ for the
pair potential. Fig.\ref{fig:f-wave} shows that this condition is satisfied for the
zigzag edge, i.e. $\Delta(k_x,-k_y)=-\Delta(k_x,k_y)$. On the other hand,
the phase difference is zero for the armchair edge, i.e. $\Delta(-k_x,k_y)=\Delta(k_x,k_y)$. Thus 
we do not expect zero-energy bound states in the latter case. The condition that there
are no trajectories with a $ \pi $-phase difference is only satisfied
for the armchair edge.  

This conclusion is consistent with the results of tight binding
model.\cite{nakada,ezawa} 
Also, similar conclusion has been recently given by the approach of
$\bm{k\cdot p}$ equation 
by Akhmerov.\cite{akhmerov}
The electronic properties of nanoribbons with clean edge based on the
$\bm{k\cdot p}$ equation can be found in ref.\cite{Phd,luis}.

Let us now turn to the original two-sublattice representation and consider the
continuum limit in order to analyze possible surface bound states. 
For zigzag edges the $x$-axis is taken to lie parallel to the edge and the
$y$-axis parallel to the normal vector. Along the edge we choose the momentum $ k_x $. 
Expanding around the $ K_+ $-point we separate the fast oscillating part by introducing
$ p_y = k_x - 2\pi/\sqrt{3}a$. Then, we write the wave function as
\begin{equation}
\psi_{A(B)}(\bm{r})=f_{A(B)}(y){\rm e}^{ik_xx}{\rm e}^{i \frac{2\pi}{\sqrt{3}a}y} ,
\end{equation}
where $ f_{A(B)} (y) $ describes the slow $y$-dependence. Concentrating on the energy close
to zero we can also take the linear momentum dependence along the $y$-direction so that 
the effective Hamiltonian has the form  
\begin{equation}
\left[
\begin{array}{cc}
0 &      h_{AB}\\
h_{BA}  & 0 
\end{array}
\right]
\left(
\begin{array}{c}
f_A(y) \\
f_B(y) \\
\end{array}
\right) 
=
E
\left(
\begin{array}{c}
f_A(y) \\
f_B(y) \\
\end{array}
\right),
\end{equation}
where
\begin{equation}
h_{BA} = h_{AB}^*= -t{\rm e}^{-i\frac{\pi}{3}}\left[
D_{k_x}-1-i\frac{a}{\sqrt{3}}
\left\{\frac{D_{k_x}}{2}+1 \right\}\hat{p}_y
\right]
\end{equation}
with $D_{k_x} = 2\cos(k_xa/2)$. Now we replace $\hat{p}_y=-i\partial_y$ and 
formulate the differential equations for $ f_{A(B)} (y) $. The outer-most lattice sites on 
the zigzag edge belong to one sublattice only, say the $B$-sublattice. Then we
take the $ A $-sublattice wave function to vanish $f_{A}(y)=0$. For the zero-energy $E=0$
we analyze the equation for $ f_B $ which reads now
\begin{equation}
\partial_y f_B(y) =
 -\frac{\sqrt{3}}{a}\frac{D_k-1}{1+\frac{D_k}{2}}f_B(y)
\equiv \alpha_B(k_x) f_B(y) \; .
\end{equation}
This equation has an exponentially decaying bound-state solution 
$f_B(y) = C{\rm e}^{\alpha_By}$ whose existence condition 
is $\alpha_B(k_x)>0$, resulting in 
$|k_x|\ge 2\pi/3$. As $k_x$ approaches $2\pi/3$ the bound state
extends deeper into the bulk, because its extension is given by
\begin{equation}
\ell(k_x) = \frac{a}{\sqrt{3}}
\frac{1+\cos(k_xa/2)}{1-2\cos(k_xa/2)}.
\end{equation}

Analogous conditions can be found for other related edges, such as the bearded 
edge where $f_{B}(y)=0$. Then we obtain zero-energy bound states for $\alpha_A = - \alpha_B > 0$
which leads to the condition $|k_x|\le 2\pi/3$ consistent with numerical calculations
for the tight-binding model.

Now we turn to the armchair edge which extends along the $y$-axis. 
Both $\bm{K_+}$ and
$\bm{K_-}$ are projected on the same point in $k_y$-momentum space.
Analogous to the zigzag edge we expand the wavefunction around one of
the two points, say 
$\bm{K_-}=\frac{2\pi}{a}\left(\frac{2}{3},0\right)$ and write
\begin{equation}
\psi_{A(B)}(\bm{r})=f_{A(B)}(x){\rm e}^{ik_yy}{\rm e}^{i\frac{4\pi}{3a}x}.
\end{equation}
where we extract again the fast oscillating part along the $x$-axis normal to the edge. 
We linearize again the spectrum close to zero energy and take 
$p_x=k_x-4\pi/3a$.
Then, for $f_{A(B)}$, the Hamiltonian has the form
\begin{equation}
\left[
\begin{array}{cc}
0 &      h_{AB}\\
h_{BA}  & 0 
\end{array}
\right]
\left(
\begin{array}{c}
f_A(x) \\
f_B(x) \\
\end{array}
\right) 
=
E
\left(
\begin{array}{c}
f_A(x) \\
f_B(x) \\
\end{array}
\right),
\end{equation}
where
\begin{equation}
h_{BA} = h_{AB}^*= -t{\rm e}^{-ik_ya/6}\left[
{\rm e}^{-ik_ya/2}-1-\frac{a}{2}\hat{p}_x
\right].
\end{equation}
where we redefined the unit length as $a\equiv\sqrt{3}a$ which
corresponding to the length of the translational vector for armchair
ribbons. Using  $\hat{p}_y=-i\partial_y$ and making the ansatz 
\begin{equation}
\left(
\begin{array}{c}
f_A(x) \\
f_B(x)
\end{array}
\right)
\sim
\left(
\begin{array}{c}
\phi_A \\
\phi_B
\end{array}
\right){\rm e}^{\lambda x} \; .
\end{equation}
The boundary condition for the armchair edge require 
$\left(\phi_A,\phi_B\right)\neq 0$ as we will see. 
For $E=0$, the $\lambda$ has two solutions 
\begin{eqnarray}
\begin{array}{cl}
\lambda_\pm & =
 -i\frac{4}{a}\sin^2\frac{ka}{4}\pm\frac{2}{a}\sin\frac{ka}{2} \\
& \equiv -i \eta \pm \zeta.
\end{array}
\end{eqnarray}
and leads to the wavefunction,
\begin{equation}
\left(
\begin{array}{c}
f_A(x) \\
f_B(x)
\end{array}
\right)
=
\left(
\begin{array}{c}
\phi^+_A \\
\phi^+_B
\end{array}
\right){\rm e}^{\lambda_+ x}
+
\left(
\begin{array}{c}
\phi^-_A \\
\phi^-_B
\end{array}
\right){\rm e}^{\lambda_- x},
\end{equation}
the coefficients are determined to satisfy the condition that
the dimensionless electronic current normal to the edge vanishes,
i.e. $j_x = 0$ with
\begin{eqnarray}
\begin{array}{cl}
j_x = & \Psi^\dagger\sigma_x\Psi\\
= & f^\ast_B(x)f_A(x) + f^\ast_A(x)f_B(x).
\end{array}
\end{eqnarray}
This leads to the solution
\begin{equation}
\left(
\begin{array}{c}
\psi_A\\
\psi_B
\end{array}
\right) 
=
\frac{1}{\sqrt{2}}
\left(
\begin{array}{c}
i\\
1
\end{array}
\right){\rm e}^{ik_yy}{\rm e}^{\frac{4\pi}{3a}x}
{\rm e}^{-i\eta x}\cosh\zeta x.
\end{equation}
It is easy to see by inserting this wavefunction into the Schr\"odinger equation that the only
zero-energy states is obtained for $ k_y = 0 $ for which $ \eta = 0 $. Thus no zero-energy bound state
exists in the case of the armchair edge.

\subsection{Spectrum and valley structure of other edges}

We now analyze the energy band structure of 
graphene nanoribbons with general edge structures based on 
the tight-binding model. We will show that the appearance of 
flat bands is a general features (apart from armchair ribbons) and
yield a two-valley structure with chiral edge states similar to the case
of zigzag edges. 

\subsubsection{Bearded Edge and Cove Edge}

We start our discussion with the most simple  edge shapes having
translational symmetry along the zigzag axis, called {\it bearded}
and {\it cove}.
Although these two edges look rather artificial compared with the zigzag edge,
they are interesting because they show very much the same non-bonding edge
localization as the standard zigzag edge. 
\begin{figure}
\includegraphics[width=0.7\linewidth]{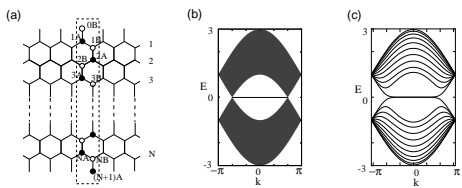}
\caption{
(a) The structure of a graphite ribbon with two bearded edges.
(b) The band structure of a semi-infinite graphite sheet
with a bearded edge.
(c) The band structure of bearded ribbon for $N=10$.
}
\label{fig2:hige2}
\end{figure}
A bearded edge is derived from a zigzag edge by adding single
$\pi$-electron hopping bonds on each  
boundary site as shown in Fig.\ref{fig2:hige2}(a).
This type of edge was first studied by Klein.\cite{klein}
In Fig.\ref{fig2:hige2}(b), the band structure of a semi-infinite
graphite sheet with a bearded edge is shown.
Interestingly, a partial flat band appears
in the region of $|k|\le 2\pi/3$, which is the
opposite condition in the semi-infinite graphite sheet with 
a zigzag edge, as we had anticipated by our continuum approximation above. 

\begin{figure}
\includegraphics[width=0.55\linewidth]{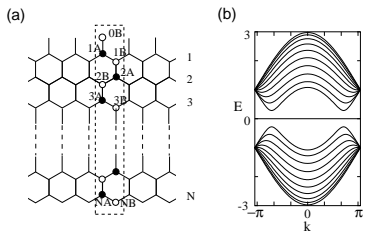}
\caption{
(a) The structure of a graphite ribbon with a zigzag and a bearded edge
and (b) the band structure for $N=10$.}
\label{fig2:hige1}
\end{figure}

It is interesting to consider a ribbon having one edge of zigzag and the other of bearded
shape as shown in Fig.\ref{fig2:hige1}(a).
Because for  this ribbon $|N_{\rm A} - N_{\rm B}| = 1$, 
where $N_{\rm A}(N_{\rm B})$ means
the number of sites belonging to the A(B)-sublattice,
there is a flat band at $E=0$ all over the 1st BZ, as
shown in Fig.\ref{fig2:hige1}(b).
The analytic solution of this flat band can be easily 
understood by the combination of two edge states for zigzag and
bearded edges.
In the region of $|k| < 2\pi/3$, the electrons are localized at 
the bearded edge, and
in the region of $|k| > 2\pi/3$, the electrons are localized at 
the zigzag edge. 
At $k=\pm 2\pi/3$, the wavefunctions extend over the whole ribbon width.
It should be noted that this ribbon is insulating at half-filling because 
the flat band has no dispersion and, thus, cannot carry currents. Moreover,
there is an energy gap between the flat band
and next subbands.

\begin{figure}
\includegraphics[width=0.8\linewidth]{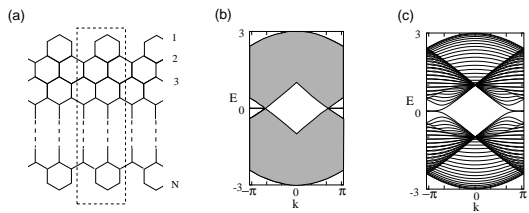}
\caption{
(a) The structure of a graphite ribbon with two cove edges. 
(b) The band structure of a semi-infinite graphite sheet
with a cove edge.
(c) The band structure of bearded ribbon for $N=10$.
}
\label{fig2:cove}
\end{figure}

Cove edge is a zigzag edge with additional hexagon rings attached.
A graphene ribbon with two cove edges is shown in
Fig.\ref{fig2:cove}(a).
In Fig.\ref{fig2:cove}(b), the band structure of a semi-infinite
graphite sheet with a cove edge is shown.
This case also provides a partly flat band 
in the region of $|k|\le 2\pi/3$ like the zigzag edge. 

Interestingly both bearded and cove graphene ribbons posses two well-separated
valleys in momentum space and chiral modes in both due to the partially flat bands.
Thus in both case PCC are realized as long as the impurity potential is long-ranged.

\subsubsection{General ribbons with $|N_A-N_B|=0$}

We extend our analysis to the electronic spectrum of nanoribbons
for which the ribbon axis is tilted with respect to the zigzag axis and
keep the balance between $A$- and $B$-sublattice sites. 
\begin{figure*}
\includegraphics[width=0.9\linewidth]{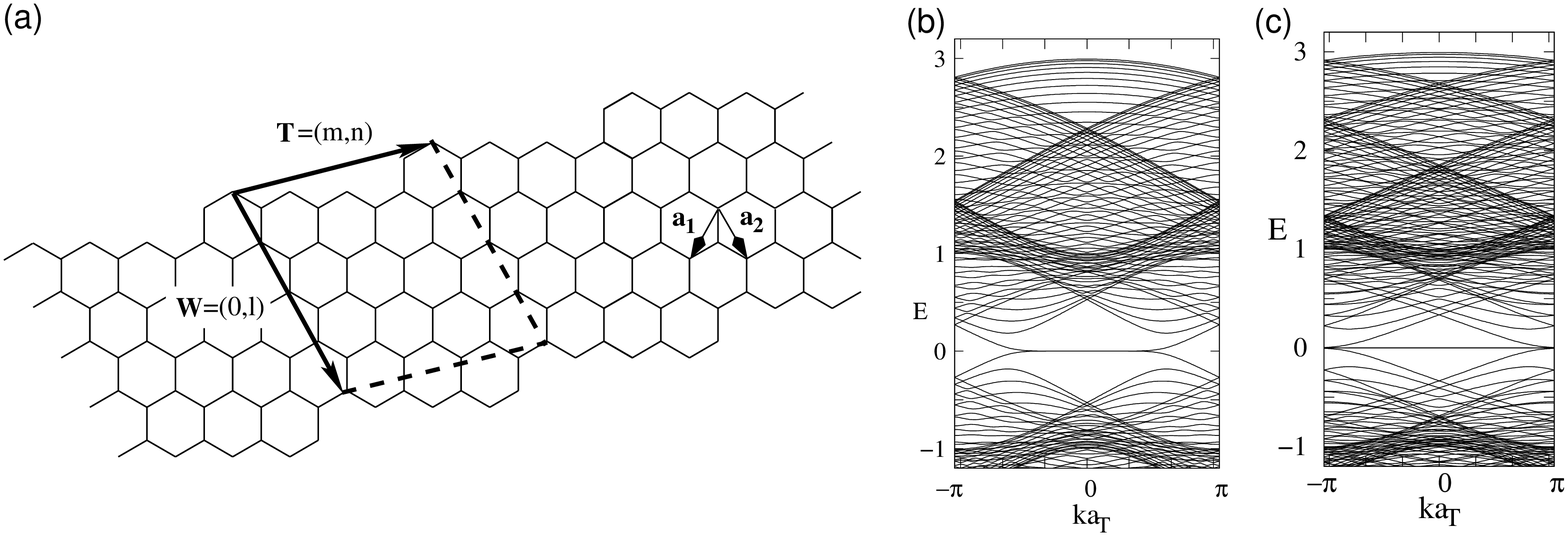}
\caption{(a) The primitive vectors for nanoribbon with the general edge
 structures. The translational vector is defined as
 $\bm{T}=(m,n)=m\bm{a_1}+n\bm{a_2}$, and
the ribbon width is defined by the vector $\bm{W}=(0,l)=l\bm{a_2}$.
The number of carbon atoms in the unit cell is $2(l+1)m$.
The corresponding energy band structures of $\bm{W}=(0,20)$ for
(b) $\bm{T}=(-4,3)$ and (c) $\bm{T}=(-6,5)$. Here $a_T$ is
the effective lattice constant which is given as $|\bm{T}|$. 
}
\label{fig:cord}
\end{figure*}
In Fig.\ref{fig:cord}(a), we show the definitions of
coordinates and primitive vectors which specifies the
geometry of the ribbon. For this purpose we introduce the two
vectors, $\bm{T} = (m,n) = m\bm{a_1}+n\bm{a_2}$ 
and $\bm{W} = (0,l)=l\bm{a_2}$,
where $l,m,n$ are integers.
The pure zigzag ribbon corresponds to $ m = - n $ and the pure armchair
edge is given by $ m =n $. 

Fig.\ref{fig:cord}(b) and (c) show the energy band structures of ribbons with
the general edge structures of $\bm{W}=(0,20)$ and 
(b) $\bm{T}=(-4,3)$ and (c) $\bm{T}=(-6,5)$ are shown. 
As we expected, the partially flat bands due to localized edge modes appear 
which break the balance between left- and right-going modes 
in the two valleys. Both examples are rather close to the zigzag edge so that the
two valleys are well separated. In this case PCC can appear.
If the geometry of the ribbons deviates more strongly from the zigzag condition, 
the valley structure will become less favorable for creating a PCC, as the
momentum difference between valleys shrinks. 
It is important to note that the extended
unit cell along these generalized ribbons reduces the valley separation drastically through
Brillouin zone folding. The length scale is the new effective lattice
constant $ a_T $ along the ribbon.　
Under these circumstances the condition for long-ranged impurity
potentials is more stringent, 
$d $ being larger than $a_T $ and not $ a $. 　

\section{Universality class}
According to random matrix theory, ordinary disordered quantum wires are classified into
the standard universality classes, orthogonal, unitary and symplectic.
The universality classes describe transport properties
which are independent of the microscopic details of disordered wires.
These classes can be specified by time-reversal and spin rotation
symmetry(Table \ref{universalityclass}).
The orthogonal class consists of systems having both time-reversal and
spin-rotation symmetries, while the unitary class is characterized by
the absence of time-reversal symmetry.
The systems having time-reversal symmetry without spin-rotation symmetry
belong to the symplectic class.
These universality classes have been believed to
inevitably cause the Anderson localization
although typical behaviors are different from class to class.
\begin{table}
\caption{Universality class}
\begin{tabular}{lcc}
Universality & TRI & SRI  \\ \hline
Orthogonal   & Yes  & Yes    \\
Unitary      & No   & irrelevant    \\
Symplectic   & Yes  & No   \\
\end{tabular}
\label{universalityclass}
\end{table}

Recently, the presence of one perfectly conducting channel
has been found in disordered metallic carbon nanotubes with LRI.\cite{suzuura}
The PCC in this system originates from the 
skew-symmetry of the reflection matrix, 
$^t\bm{r}= -{\bm r}$,\cite{suzuura}
which is special to the symplectic symmetry
with odd number of channels.
The electronic transport properties such system
has been studied on the basis of the random matrix theory.\cite{takane1, takane2}
On the other hand, zigzag ribbons without inter-valley scattering are
not in the symplectic class, since they break TRS in a special way. 
The decisive feature for a perfectly conducting channel is
the presence of one excess mode in each valley as discussed in the
previous section.

In view of this classification we find that the universality class of
the disordered zigzag ribbon 
with long-ranged impurity potential (no inter-valley scattering) is the {\it unitary} class 
(no TRS). On the other hand, for short-range impurity potentials 
with inter-valley scattering the disordered ribbon belongs to the {\it orthogonal} class 
(with overall TRS). This classification is 
compatible with the magnetic field dependence of the localization length
$\xi$ as shown in Fig.\ref{fig:xi}.
Consequently we can observe a crossover between two universality classes when we change
the impurity range continuously. 

\begin{figure}
\includegraphics[width=0.7\linewidth]{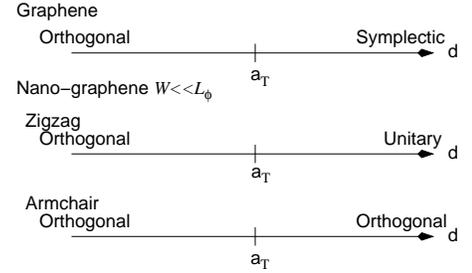}
\caption{Summary concerning for the universality crossover. With
 increasing the range of the impurity potential, graphene is known to be
the orthogonal for SRIs and the symplectic class for LRIs. However,
zigzag nanoribbons are unitary class for SRIs. Armchair ribbons are
irrelevant for the range of the impurity. $L_\phi$ is the phase
 coherence length. $W$ is the width of graphene ribbons.
}
\label{fig:crossover}
\end{figure}

Analogous symmetry considerations can be applied to armchair ribbons. In this case
the two valleys merge into a single one at $k=0$. TRS is conserved
irrespective of the impurity potential range, if there is no magnetic field.
Consequently, disordered armchair ribbons belong always to the
orthogonal class and do not provide a perfectly conducting channel. 
In view of the fact that graphene is known to be symplectic (orthogonal)
for LRI (SRI),\cite{suzuura.prl} it is 
quite intriguing to realize that the edges influence the universality
class, as long as the phase coherence length is larger than the system
characteristic size of the nanographene system. In the Fig.\ref{fig:crossover}, the summary of
the argument is visualized. 
In the nano-graphene ribbons with general edges, 
since the two Dirac points are separated in the momentum space
with the order of $|1/\bm{T}|$, 
the characteristic length causing the crossover is $\sim |\bm{T}|$.
Thus the PPC is expected to appear even in generic nanoribbons
if the sample has only very slowly varying charge potential and
the intervalley scattering is suppressed.

\section{Conclusion}
The unusual energy dispersion due to their edge states gives rise to 
the unique property of zigzag ribbons. Concerning transport properties for disordered
systems the most important consequence is the presence of a perfectly
conducting channel, i.e. the absence of Anderson localization which is
believed to inevitably occur in the one-dimensional electron system.
The origin of this effect lies in the single-valley transport which is
dominated by a chiral mode. On the other hand, large momentum transfer
through impurities with short-range potentials involves both valleys,
destroying this effect and leading to usual Anderson localization.  The
obvious relation of the chiral mode with time reversal symmetry leads
to the classification into the unitary and orthogonal class 
depending on the range of impurity potential. Since the 
inter-valley scattering is weak in the experiments of graphene, we may assume that these
conditions may be realized also for ribbons. Naturally defects  in the 
ribbon edges and vacancies would be rather harmful for the experiment
making this type of experiment very challenging.\cite{prl,prb}

\section*{ACKNOWLEDGEMENT}
We thank T. Enoki, K. Kusakabe for stimulating discussions. 
K. W. acknowledges the financial support by a 
Grant-in-Aid for Young Scientists (B) (No. 19710082)
from the Ministry of Education,
Culture, Sports, Science and Technology (MEXT).
This work was financially 
supported by the Swiss Nationalfonds through Centre for Theoretical Studies of
ETH Z\"{u}rich and the NCCR MaNEP, also supported by a Grand-in-Aid for
Scientific Research (B) and (C)  
from the Japan Society for the Promotion of Science (No. 19310094,
No. 16540291).
The numerical calculation was performed on the Grid/Cluster
Computing System and HITACHI SR11000 at Hiroshima University.


\begin{thebibliography}{99}
\bibitem{note.bc}
Note two virtual zigzag chains on both sides are necessary
for the boundary condition on zigzag ribbon of width $N$.
\end{thebibliography}
\end{document}